\documentclass[pdftex,twocolumn,epjc3]{svjour3}          
\usepackage[utf8]{inputenc}
\usepackage{amsmath,amssymb,graphicx,textcomp,enumerate,alltt,xspace,multirow,xcolor,mathrsfs}
\usepackage{subfig}
\usepackage{fancyvrb}
\usepackage{mathtools}

\newcommand\POWHEGBOX{{\tt POWHEG-BOX}}
\newcommand\FONLL{{\tt FONLL}}
\newcommand\POWHEG{{\tt POWHEG}}

\newcommand\bbfourl{{\tt bb4l}\xspace}

\newcommand\hvq{{\tt hvq}}

\newcommand{\mathd}{\mathrm{d}}

\newcommand\sss{\mathchoice%
{\displaystyle}%
{\scriptstyle}%
{\scriptscriptstyle}%
{\scriptscriptstyle}%
}

\newcommand{\bj}{\ensuremath{{j_{\rm \sss B}}\xspace}}
\newcommand{\bbj}{\ensuremath{j_{\rm \sss {\bar B}}}\xspace}

\newcommand{\hvqash}{{\tt nol}}
\newcommand{\hvqasl}{{\tt asl}}
\newcommand{\hvqasldef}{{\tt def}}
\newcommand{\hvqaslalt}{{\tt alt}}

\def\beq{\begin{equation}}
\def\beqn{\begin{eqnarray}}
\def\eeq{\end{equation}}
\def\eeqn{\end{eqnarray}}

\def\({\left(} 
\def\){\right)} 
 
\newcommand     \MSB            {\ifmmode {\overline{\rm MS}} \else
                                 $\overline{\rm MS}$\fi}

\newcommand\as{\alpha_{\sss\rm S}}

\newcommand\pT{\ensuremath{p_{\rm T}}}

\newcount\minutes 
\newcount\scratch 
\def\timestamp{%
\scratch=\time 
\divide\scratch by 60 
\edef\hours{\the\scratch} 
\multiply\scratch by 60 
\minutes=\time 
\advance\minutes by -\scratch 
---$\,$\hours:\null 
\ifnum\minutes< 10 0\fi 
\the\minutes}

\definecolor{mygray}{gray}{0.5}

\newcommand\blankout[1]{}

\RequirePackage[T1]{fontenc}

\smartqed  

\RequirePackage{graphicx}
\RequirePackage{mathptmx}      
\RequirePackage[numbers,sort&compress]{natbib}
\RequirePackage[colorlinks,citecolor=blue,urlcolor=blue,linkcolor=blue]{hyperref}

\journalname{Eur. Phys. J. C}

\usepackage{listings}
\begin{document}

\title{Heavy quark radiation in NLO+PS POWHEG generators}

\author{Luca Buonocore\thanksref{e1,addr1}
        \and
        Paolo Nason\thanksref{e2,addr2}
        \and
        Francesco Tramontano\thanksref{e3,addr1}
}

\thankstext{e1}{e-mail: luca.buonocore@na.infn.it}
\thankstext{e2}{e-mail: paolo.nason@mib.infn.it}
\thankstext{e3}{e-mail: francesco.tramontano@na.infn.it}

\institute{Universit\`a di Napoli ``Federico II'' and INFN, Sezione di Napoli,
  Complesso di Monte Sant'Angelo, Via Cintia, 80126 Napoli, Italy\label{addr1}
          \and
           Theoretical Physics Department, CERN, Geneve, Switzerland, and INFN, Sezione di Milano-Bicocca, Milano, Italy\label{addr2}
}

\date{CERN-TH-2017-265}

\maketitle

\begin{abstract}
  In this paper we deal with radiation from heavy quarks in the context
  of next-to-leading order calculations matched to parton shower generators.
  A new algorithm for radiation from massive quarks is presented that has
  considerable advantages over the one previously employed.
  We implement the algorithm in the framework of the \POWHEGBOX{},
  and compare it with the previous one in the case of
  the \hvq{} generator for bottom production in hadronic collisions,
  and in the case of the \bbfourl{} generator
  for top production and decay.
\end{abstract}

\section{Introduction}
The production and detection of bottom quarks play an important r\^ole in various
contexts in LHC physics.
Letting aside the very abundant direct production, that is exploited for flavour physics
studies, bottom is used to identify top particles and to study their properties. Furthermore,
it is the dominant decay mode of the Higgs boson, that can be used to study processes
as the associate $HV$ production \cite{Aaboud:2017xsd,CMS:2017tkf}
and the large transverse momentum production \cite{CMS:2017cbv}.
In searches for physics beyond the Standard Model, bottom also appears often produced in
association with new-physics objects.

Having a mass much larger than the typical hadronic scales, bottom quark production is
calculable in perturbative QCD. In cases when the transverse momentum involved in the
production is large compared to its mass, as, for example, in high-energy $e^+e^-$
annihilation, or in production at large transverse momentum in hadronic collisions,
bottom can behave as a light parton, and give rise to a hadronic jet.
Techniques for dealing with these regimes have been developed in the past~\cite{Cacciari:1998it},
and have
been applied to the LHC case~\cite{Cacciari:2012ny}. They allow for the computation
of the transverse momentum spectrum of promptly produced $b$ quarks at next-to-leading
order in QCD, including the resummation of large logarithms of the ratio of the
transverse momentum over the bottom mass up to next-to-leading-logarithmic accuracy.
These large logarithms can arise both from initial state radiation,
when, for instance, an incoming gluon splits into a $b\bar{b}$ pair, with one of the $b$
undergoing a large-momentum-transfer collision with a parton from the target, and
from final state radiation. In the last case, an outgoing gluon can split into
a $b\bar{b}$ pair, or a directly produced $b$ quark can emit a collinear gluon.

The large transverse momentum regime is treated consistently at the leading logarithmic level
in parton shower generators. At the most basic level,
heavy flavours are treated as light flavours, but with
a shower cut-off scale of the order of the heavy quark mass.
However, considerable work has been performed to better account for mass effects.
In some generators, this is achieved by suitable modifications of the splitting kinematics
and splitting kernels~\cite{Sjostrand:2014zea,Bellm:2017bvx}.
The Sherpa dipole shower \cite{Schumann:2007mg} makes use of the
Catani-Seymour dipoles for massive quarks~\cite{Catani:2002hc}.
The Catani-Seymour formalism is also used in the DIRE shower~\cite{Hoche:2015sya}.
In ref.~\cite{GehrmannDeRidder:2011dm} a final state dipole-antenna shower
for massive fermions is proposed, based upon the corresponding
antenna subtraction formalism of ref.~\cite{GehrmannDeRidder:2009fz}.

In next-to-leading order (NLO) calculations matched to Shower
generators (NLO+PS) for heavy flavour
production~\cite{Frixione:2003ei,Frixione:2007nw}, one generally
treats the heavy flavour as being very heavy.  The heavy quark mass
thus acts as a cut-off on collinear singularities, that are thus not
resummed. This approach has in fact proven to be quite viable in heavy
flavour production even at relatively large momentum
transfer~\cite{Cacciari:2012ny}. Consider, for example, heavy quark
pair production in a \POWHEG{} framework.  By neglecting collinear
singularities from heavy quarks, the only singular region that we have
to consider has to do with initial state radiation involving only
light partons. Since the \POWHEG{} procedure guarantees that the
matrix elements are given correctly for up to one hard radiation,
gluon splitting, flavour excitation and radiation from the heavy
flavour are included, so that the logarithmically enhanced terms are
correctly reproduced at first order.  Higher order leading logarithms,
however, are not treated correctly.  In particular, there are reasons
to give an adequate treatment to final state radiation from a high
transverse momentum bottom quark. In fact, this radiation process is
intimately related to the physics of the bottom fragmentation
function, and may have important effects in processes of considerable
interest, like for example in top decay.

In the \POWHEGBOX{} framework, a facility for the treatment of
collinear radiation from a heavy particle was set up in
ref.~\cite{Barze:2012tt}, in the framework of electroweak corrections
to $W$ production. In that context, the purpose was to deal with mass
effects in the final state radiation of photon from the lepton in $W$
decays. The same framework is also appropriate
for describing radiation from heavy quarks. In particular, it was
adopted in refs.~\cite{Campbell:2014kua}, where the \verb!ttb_NLO_dec!
generator was introduced, and in refs.~\cite{Jezo:2016ujg}
for the \verb!b_bbar_4l! generator, for dealing with radiation
from bottom quarks in top decays.

The purpose of the present paper is twofold: we present a new algorithm
for radiation from a heavy quark, that has proven superior to the old one;
furthermore we perform a thorough investigation of the behaviour of this
component of the \POWHEG{} generator, also by comparing the two methods,
both in the framework of bottom quarks generated in top decay, and
in inclusive bottom quark pair production. In the last case,
such a study was never carried out.

The paper is organized as follows: in section~\ref{sec:descr} we
describe the new algorithm, in section~\ref{sec:pheno} we illustrate
our phenomenological studies, and in section~\ref{sec:conc} we give our
conclusions.

\section{Description of the new algorithm}\label{sec:descr}
\blankout{
\begin{itemize}
\item Tentatively, we assume that the jacobian singularity
  is really there and is of square root type.
\item We should show that in fact it is weak, and does not
  prevent doing the integrations (that can also be done by importance
  sampling) end the generation of radiation.
\end{itemize}
}
\subsection{The \POWHEG{} mapping for the massive emitter case}
\begin{figure}
  \centering
  \includegraphics[scale=0.35]{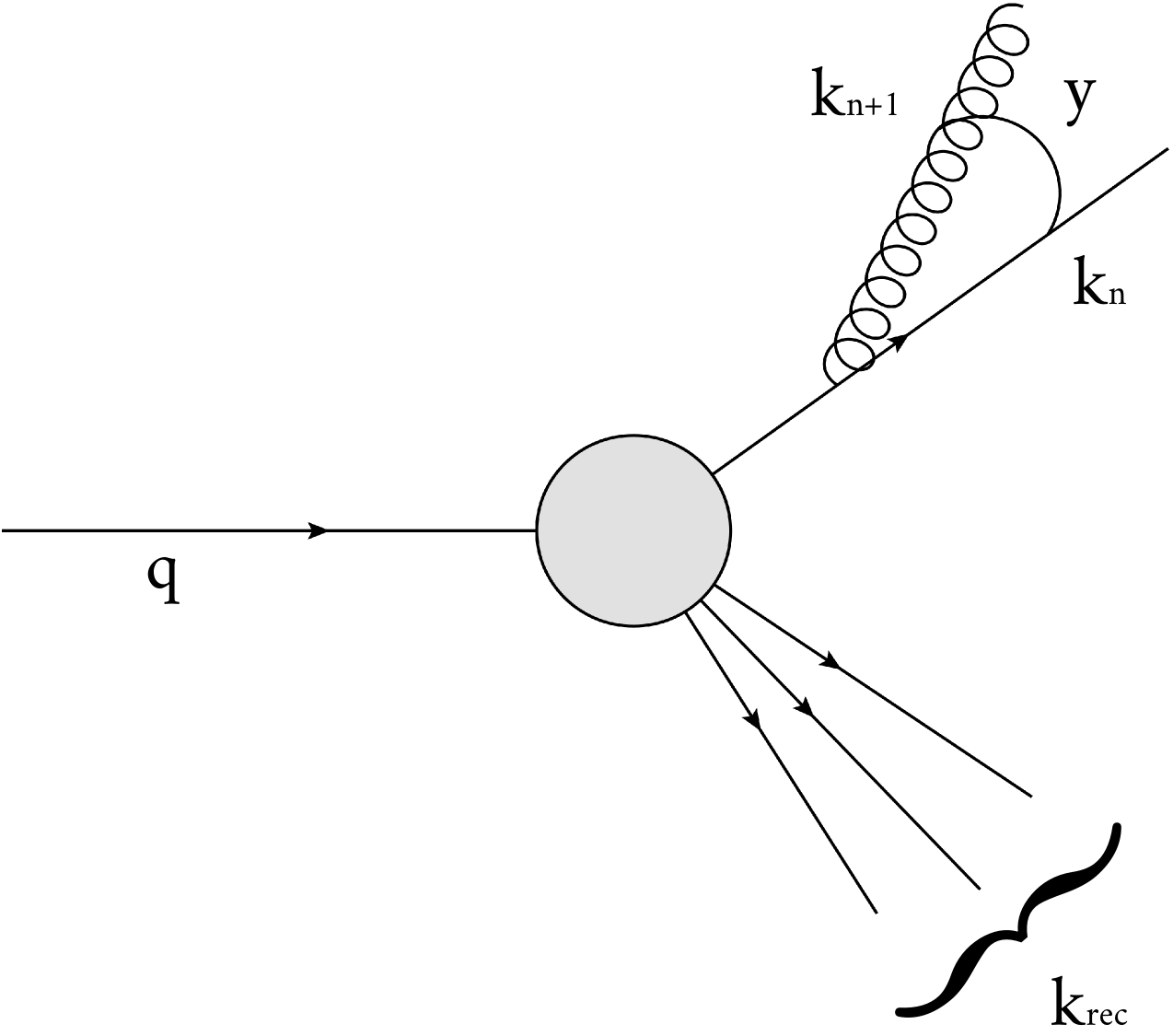}
  \caption{Kinematics for a real configuration: $k_n$ is the massive emitter, $k_{n+1}$ is the radiated parton. $y$ denotes the cosine of the angle between the two tri-vectors.}\label{fig::kinematics}
\end{figure}

Let us assume for definiteness to deal with a scattering process involving $n$ partons in the final state at lowest order in perturbation theory. The generic point in the corresponding Born phase space, referred also as Born configuration, will be denoted with barred momenta
\begin{equation}
  \overline{\Phi}_n = \{\overline{k}_1, \dots, \overline{k}_n\}.
\end{equation}
The corresponding phase space volume element is given by
\begin{equation}
  \mathd\overline{\Phi}_n= \prod_{i=1}^n\frac{\mathd^3\vec{\overline{k}}_i}{(2\pi)^32\overline{k}_i^0}(2\pi)^4\delta^{(4)}\left(q-\sum_{i=1}^n\overline{k}_i\right),
\end{equation}
where $q$ is the total incoming 4-momentum.\footnote{The system we are
  considering can be either the full final state, or the system of decay products of a resonance,
  according to the origin of the heavy quark.}
At Next-to-Leading order (NLO), one must also include processes of emission of
one more real massless extra parton, resulting in a $n+1$-body kinematics which we will denote as
\begin{equation}
  \Phi_n = \{k_1, \dots, k_{n+1}\}.
\end{equation}
The singular regions of the real phase space
are separated by means of suitable projection operators; in each of them, the radiated parton phase space is parametrised in terms of the FKS variables \cite{Frixione:1995ms} (the notations $\vec{p}$ and $\underline{p}$ for a generic momentum $p$ denote the tri-impulse and its modulus respectively) 
\begin{equation}
  \xi=\frac{2\underline{k}_{n+1}}{q^0}, \quad y=\frac{\vec{k}_{n}\cdot\vec{k}_{n+1}}{\underline{k}_n\underline{k}_{n+1}},
\end{equation}
as shown in Fig.~\ref{fig::kinematics}, where we have assumed that the emitter and the FKS partons are respectively the $n$-th and the $n+1$-th parton. The rescaled energy $\xi$ is related to the soft limit ($\xi \to 0$), and the variable $y$ to the collinear one ($y\to \pm 1$). The definition of the azimuthal angle $\phi$, in the \POWHEG{} framework, departs
from the standard FKS definition. It is taken as the polar angle of the splitting around the
axis parallel to the momentum of the recoil system, in the rest frame where $q=(q^0,\vec{0})$.
\par In what follows, we will construct a one-to-one map from a real configuration with radiation variables $(\xi,y,\phi)$ into a Born one. This leads to a factorisation of the real phase space in term of Born and radiation variables. 
\par The mapping can be reduced to the case of the map from a 3-body phase space into a 2-body one. Inserting into the $(n+1)$-body phase space volume element
the identities
\begin{equation}
  1=\int \mathd^4k_{\text{rec}}\delta^{(4)}\left(k_{\text{rec}}-\sum_{i=1}^{n-1}k_i\right)
\end{equation}
and
\begin{equation}
  1=\int \mathd M^2_{\text{rec}}\delta(M^2_{\text{rec}}-k_{rec}^2),
\end{equation}
the phase space is decomposed into a chain of two consecutive processes.
With reference to Fig.~\ref{fig::kinematics}, they are: the decay of a particle with momentum $q$ into the 3-body system formed by the emitter $k_n$, the FKS-parton $k_{n+1}$ and the ``recoil'' system, with momentum and invariant mass
\begin{equation}
  k_{\text{rec}}= \sum_{i=1}^{n-1}k_i= q-k_n-k_{n+1}, \quad M_{\text{rec}}^2=k_{\text{rec}}^2,
\end{equation} 
followed by the decay of the latter into the other $n-1$ particles. In formula, we have
\begin{equation}\label{fac_recoil}
  \mathd\Phi_{n+1}=\mathd\Phi_3\mathd\Phi_{\text{rec}},
\end{equation}
where
\begin{equation}\label{three_body}
\begin{split}
  \mathd\Phi_3=&\frac{\mathd M^2_{\text{rec}}}{2\pi}\frac{\mathd^3 \vec{k}_n}{2k^0_n(2\pi)^3}\frac{\mathd^3 \vec{k}_{n+1}}{2k^0_{n+1}(2\pi)^3}\frac{\mathd^3 \vec{k}_{\text{rec}}}{2k^0_{\text{rec}}(2\pi)^3}\\&\times(2\pi)^4\delta^{(4)}(q-k_n-k_{n+1}-k_{\text{rec}}),
\end{split}
\end{equation}
\begin{equation}
\mathd\Phi_{\text{rec}}= \prod_{i=1}^{n-1}\frac{\mathd^3\vec{k}_{i}}{2k_i^0(2\pi)^3}(2\pi)^4\delta^{(4)}\left(k_{\text{rec}}-\sum_{i=1}^{n-1}k_i\right). 
\end{equation}
We now focus on the 3-body process; under the action of the mapping, the $k_n$ and $k_{n+1}$ partons will be replaced by a single parton with mass $m$ and momentum $\overline{k}_n$. We define
\begin{equation}
  k\equiv k_n+k_{n+1},
\end{equation}
so that
\begin{equation}
  k_{\text{rec}}=q-k \implies k^0_{\text{rec}}=q^0-k^0,\vec{k}_{\text{rec}}=-\vec{k}.
\end{equation}
We fix the transformation by demanding $\vec{\overline{k}}_n \parallel \vec{k}$. Care must be taken to ensure the conservation of energy-momentum also for the resulting Born configuration. This is accomplished by performing a boost $\Lambda$ in the direction $\vec{k}$ and defining
\begin{equation}\label{barred_emitter}
\overline{k}_n=q-\Lambda k_{\text{rec}},
\end{equation}
We determine the velocity parameter $\beta$ of the boost transformation from the mass-shell condition
\begin{equation}
  \overline{k}_n^2 = (q-\Lambda k_{\text{rec}})^2 = m^2.
\end{equation}
We get
\begin{equation}\label{beta_boost}
\begin{split}
  \beta &= \frac{-4\underline{k}_{\text{rec}}k^0_{\text{rec}}q^2}{(q^2-m^2+M_{\text{rec}}^2)^2+4\underline{k}_{\text{rec}}^2q^2} \\ &\hspace{0.5cm}+\frac{(q^2-m^2+M_{\text{rec}}^2)\sqrt{(q^2-m^2+M_{\text{rec}}^2)^2-4M_{\text{rec}}^2q^2}}{(q^2-m^2+M_{\text{rec}}^2)^2+4\underline{k}_{\text{rec}}^2q^2}.
\end{split}
\end{equation}
We define the other barred variables as
\begin{equation}
  \overline{k}_i= \Lambda k_i, \quad i=1,\dots,n-1.
\end{equation}
Their mass relations are preserved by the boost transformation and, furthermore, we have
\begin{equation}
\begin{split}
  \sum_{i=1}^{n}\overline{k}_i&= \sum_{i=1}^{n-1}\overline{k}_i+\overline{k}_{n}= q+\sum_{i=1}^{n-1}\Lambda k_{i}-\Lambda k_{\text{rec}} \\&= q+\Lambda\bigg(\sum_{i}^{n-1}k_i-k_{\text{rec}}\bigg)=q,
\end{split}
\end{equation}
which is the energy-momentum conservation for the Born configuration.
\subsection{Inverse map}
\par We now detail the construction of the inverse map, which is
what is actually needed in the applications.
Suppose that a Born event has been generated, i.e. the barred variables $\overline{k}_i$ ($i=1,\cdots,n$) are given.
Then, $M_{\text{rec}}^2$ is obtained inverting eq.~\eqref{barred_emitter}:
\begin{equation}
  M_{\text{rec}}^2= (\Lambda k_\text{rec})^2 = (q-\overline{k}_n)^2= q^2+m^2-2q^0\overline{k}_n^0.
\end{equation}
We want to attach to it a radiation described by the radiation variables $\xi$, $y$ and $\phi$.
For future convenience we introduce the largest allowed value for $\xi$
\begin{equation}\label{ximax}
  \xi_{\rm max}\equiv 1-\frac{(m+M_{\rm rec})^2}{q^2}.
\end{equation}
The energy of the radiated parton is
\begin{equation}
  k^0_{n+1}= \underline{k}_{n+1}=\frac{q^0}{2}\xi.
\end{equation}
Energy conservation requires that
\begin{equation}  \label{encons}
  q^0 = k^0_{n+1}+ \sqrt{\underline{k}_n^2+m^2}
  + \sqrt{\underline{k}_{\rm rec}^2+M_{\text{rec}}^2},
\end{equation}
where
\begin{equation}
  \underline{k}_{\rm rec}^2=\underline{k}_n^2+\underline{k}_{n+1}^2+2\underline{k}_n\underline{k}_{n+1}y.
\end{equation}
We can solve equation~(\ref{encons}) for $\underline{k}_n$ in a standard way, by bringing in turn each
single square root
on one side of the equation and squaring both sides. By doing this we actually find the
solutions of all of the following equations
\begin{equation}\label{allpmeq}
q^0 = k^0_{n+1} \pm \sqrt{\underline{k}_n^2+m^2} \pm \sqrt{\underline{k}_{\rm rec}^2+M_{\text{rec}}^2},
\end{equation}
for all possible combinations of the signs in front of the square root. The solutions are given by
\begin{equation}\label{knsolpm}
\begin{split}
  \underline{k}_{n}^{(\pm)} &= \frac{-(2\overline{k}_n^0-q^0\xi)\xi y}{(2-\xi)^2-\xi^2y^2} \\
  &\pm \frac{(2-\xi)\sqrt{(2\overline{k}_n^0-q^0\xi)^2 -m^2\xi ^2(1-y^2)-4m^2(1-\xi)}}{(2-\xi)^2-\xi^2y^2}.
\end{split}
\end{equation}
In order for them to exist, the argument of the square root must
be positive. This leads to the bound
\begin{equation}\label{knpmbound}
  (q^2-m^2+m^2y^2) \xi^2 - 4(q^0\overline{k}_n^0 -m^2) \xi + 
  4 \overline{\underline{k}}_n^2 >0,
\end{equation}
with $\overline{\underline{k}}_n^2=(\overline{k}_n^0)^2-m^2$.
Eq.\eqref{knpmbound} is satisfied if either $\xi>\xi^{(+)}(y)$ or
$\xi<\xi^{(-)}(y)$, with
\begin{eqnarray}
  \xi^{(\pm)}(y) &=&
  2\frac{\overline{k}_n^0q^0 - m^2  \pm
    m \sqrt{(q^0-\overline{k}_n^0)^2-\overline{\underline{k}}_n^2y^2}}{q^2-m^2+m^2y^2},
  \nonumber \\
  &=&
   \frac{q^2 - m^2 -M_{\rm rec}^2 \pm
     2  m \sqrt{M_{\rm rec}^2+\overline{\underline{k}}_n^2(1-y^2)}}{q^2-m^2+m^2y^2}
   \nonumber \\
  &=& \frac{4\overline{\underline{k}}_n^2}
   {q^2 - m^2 -M_{\rm rec}^2 \mp
  2  m \sqrt{M_{\rm rec}^2+\overline{\underline{k}}_n^2(1-y^2)}}.
\end{eqnarray}
The last equality follows from the fact that
\begin{equation}
\xi^{(+)}\xi^{(-)}=\frac{4\overline{\underline{k}}_n^2}{q^2-m^2+m^2y^2}.
\end{equation}
We see that $\xi^{(+)}$ is a decreasing function of $y^2$. Thus
\begin{equation}
  \xi^{(+)}(y) > \xi^{(+)}(1)=1-\frac{(m-M_{\rm rec})^2}{q^2} > \xi_{\rm max}.
\end{equation}
that is larger than
the maximum value allowed by energy conservation. Thus, the corresponding
$\underline{k}_{n}^{(\pm)}$ values should be the solutions of
one among equations~(\ref{allpmeq}) where some minus signs appear.
On the other hand,  $\xi^{(-)}(y)$ is an increasing function of $y^2$, so
\begin{equation}
  \xi^{(-)}(y) < \xi^{(-)}(1) = 1-\frac{(m+M_{\rm rec})^2}{q^2},
\end{equation}
that is perfectly acceptable. Furthermore, in the $\xi<\xi^{(-)}(y)$ case the value $\xi=0$
is allowed, that lead
to the solutions $\underline{k}_{n}^{(\pm)} = \pm \overline{k}_n^0$ satisfying
eq.~(\ref{encons}) with the correct signs of the square roots.
Since the $\underline{k}_{n}^{(\pm)}$ must always satisfy  one of the equations~(\ref{allpmeq}),
and since they are smooth function of both $\xi$ and $y$ in their allowed range
(that includes the  $\xi=0$ point), we infer by continuity that they
satisfy equation~(\ref{allpmeq}).

\begin{figure}[]
  \centering
  \includegraphics[scale=0.45]{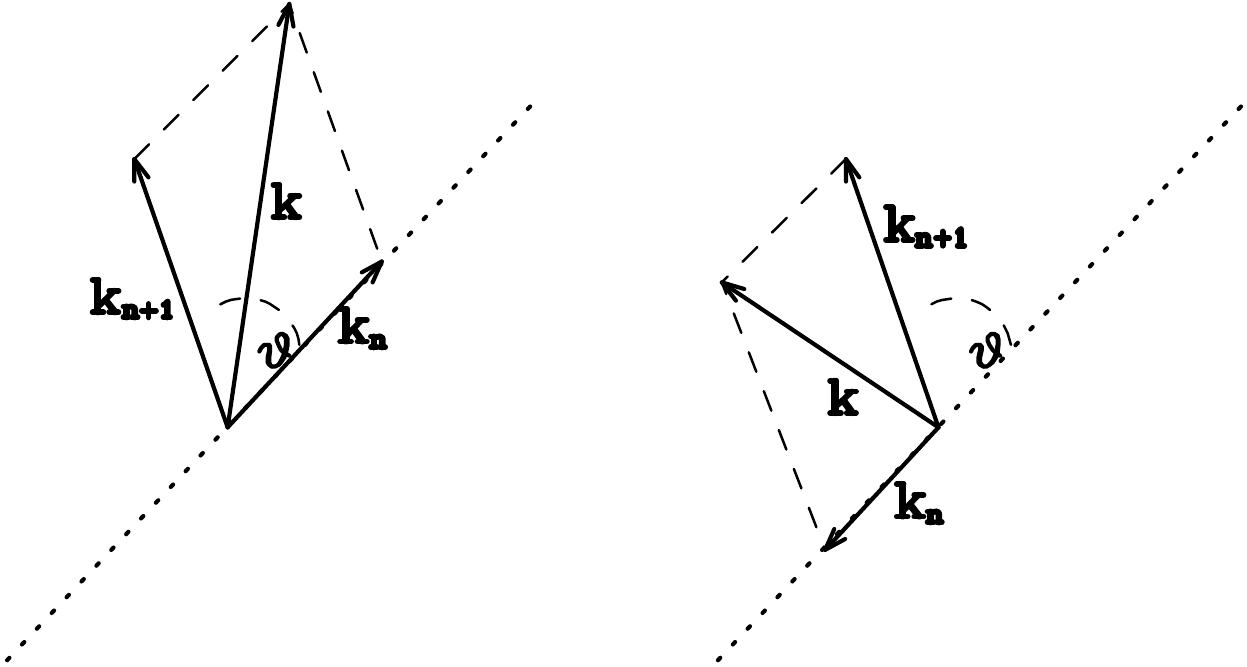}
  \caption{Kinematic reconstruction of the real emission kinematics with positive (left) and negative
  $\underline{k}_n$ values. The angle $\theta$ is fixed by $y=\cos\theta$.}
  \label{negkvalues}
\end{figure}

\begin{figure}[]
  \centering
  \includegraphics[width=0.45\textwidth]{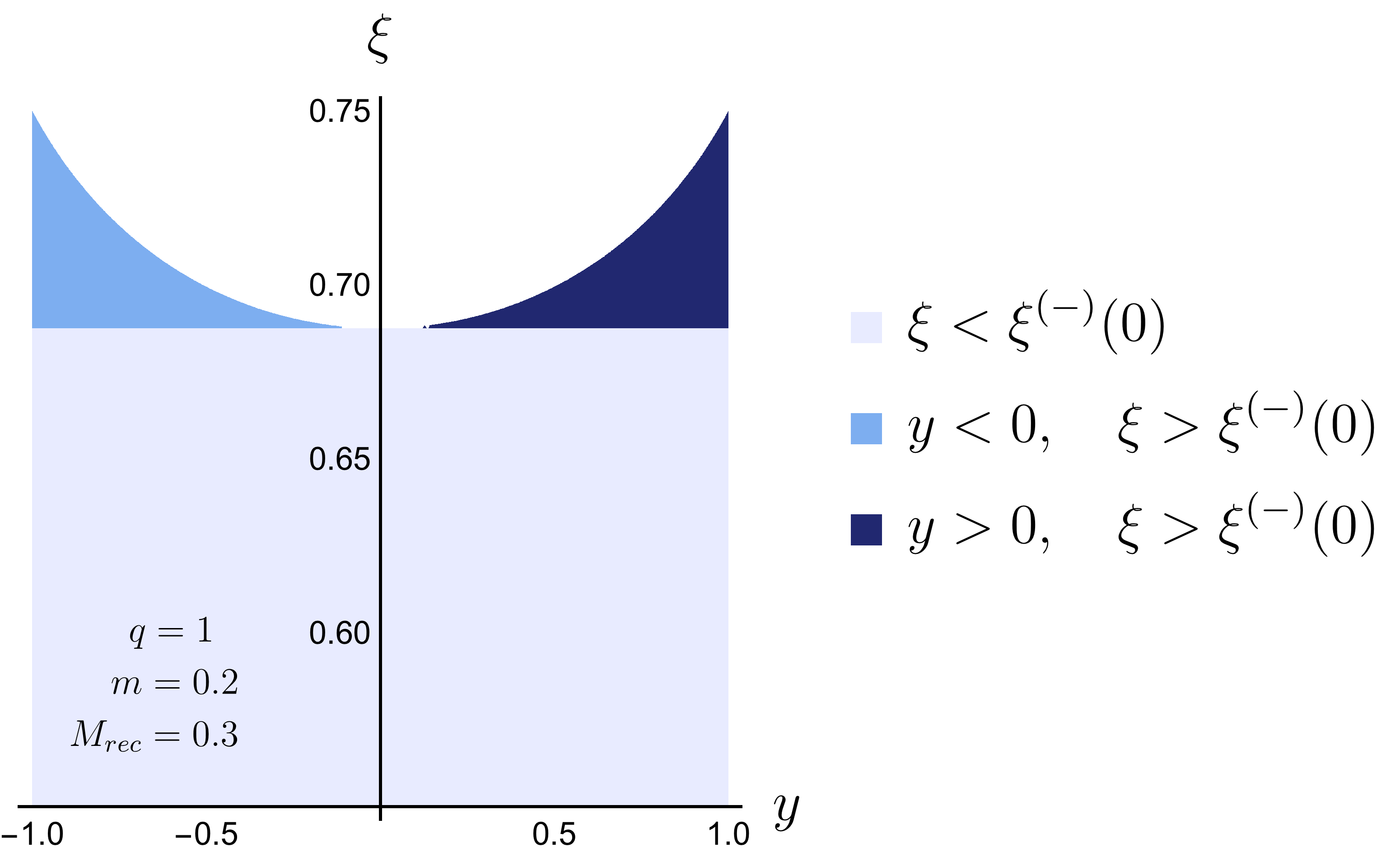}
  \caption{Plot of the physical region in the $\xi y$ plane. The shaded orange region is where $\underline{k}_n^{(+)}(\xi,y)$ is negative.
    It is physically equivalent to the (positive) $\underline{k}_n^{(-)}(\xi,-y)$ solution in the dark blue region. If we insisted upon
    considering only positive $\underline{k}_n$ solutions, the blue region would be doubly covered, and the dark blue one would not be there.}
  \label{fig::xiupperlimit}
\end{figure}

Up to now we have not imposed the positivity of $\underline{k}_n$. On the other hand, negative $\underline{k}_n$
values still have a physical interpretation, as illustrated in fig.~\ref{negkvalues}.
Thus, provided we interpret negative values of $\underline{k}_n$ according to the construction
of fig.~\ref{negkvalues}, we have two solutions
of equation~\eqref{encons}. They are however related, since
\begin{equation}
\underline{k}_n^{(+)}(\xi,y)=-\underline{k}_n^{(-)}(\xi,-y).
\end{equation}
If we pick just one of them, we have a single-value map from the underlying Born configuration and the
radiation variables $\xi$, $y$ and $\phi$ to a real emission configuration. We pick the solution
$\underline{k}_n^{(+)}(\xi,y)$, since for $m=0$ it corresponds to the usual solution
in the massless case. Unlike in the massless case, however, $\underline{k}_n^{(+)}(\xi,y)$ is not always positive:
it is negative in the region
\begin{equation}
  y > 0 \,,\quad
  \xi > \xi^{(-)}(0) = 2\frac{\overline{k}^0_n -m}{q-m}=
  \frac{(q^0-m)^2-M_{rec}^2}{q^0\,(q^0-m)}.
\end{equation}
For continuity, $\underline{k}_n^{(+)}(\xi,y)$ vanishes on the boundary line
$y>0,\;\xi=\xi^{(-)}(0)$
separating the positive and negative regions. The points lying on this curve
are degenerate and correspond to the same real configuration 
with the emitter at rest in the partonic centre-of-mass frame.
Apart from them, that constitute a set
  of zero measure, the map is well defined and bijective. The inverse map is
well defined also on the boundary line $y>0,\, \xi=\xi^{(-)}(0)$.
This means that the corresponding jacobian
vanishes on that curve. Then, the inverse map
can be safely used both for the integration of the real differential cross
section and for the generation of radiation. \\
In fig.~\ref{fig::xiupperlimit}
we display the $\xi,y$ kinematic region.
We remark that the negative  $\underline{k}_n^{(+)}(\xi,y)$ region includes neither soft
nor collinear singularities, since $\xi$ is large, and
since the angular separation of the quark and the radiated gluon is larger than $\pi/2$.
From now on we will drop the suffix $(-)$ and will use $\xi(y)$ and $\xi(0)$ instead of
$\xi^{(-)}(y)$ and $\xi^{(-)}(0)$.

\begin{figure}
  \centering
 \includegraphics[width=0.45\textwidth]{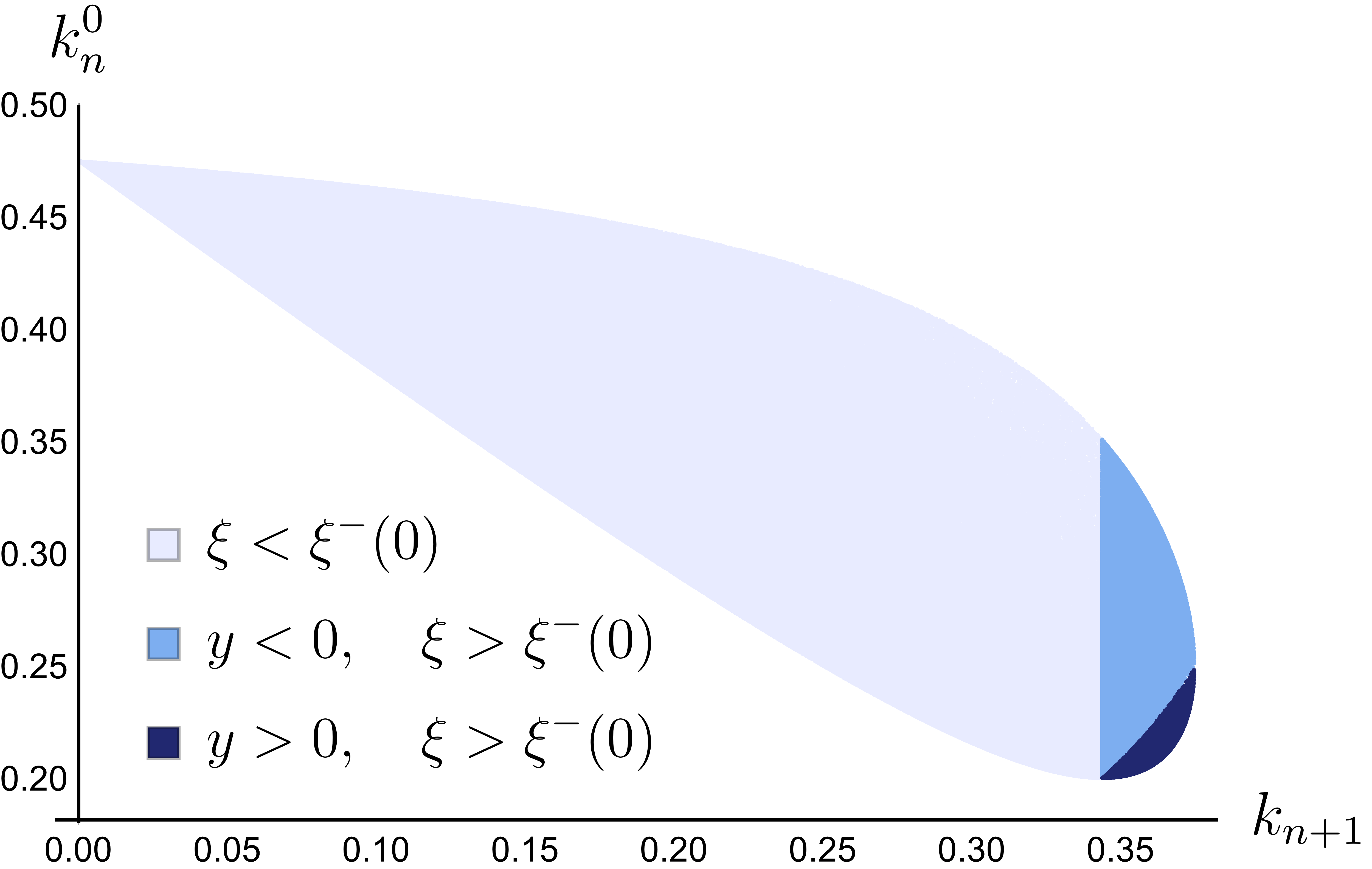}
 \caption{Dalitz plot for the three-body phase space of the system comprising
   the heavy flavour, the radiated gluon and the recoiling
   system.}\label{fig:dalitz}
\end{figure}

In fig.~\ref{fig:dalitz} we show the partition o the kinematic region
represented in the more familiar Dalitz plane.
Notice that in the massless limit the physical region in the Dalitz plot
develops an acute angle in the lower right, corner corresponding to the  
gluon being anticollinear with the $b$ quark.
Thus, the problematic region $\xi>\xi(0)$ is not a singular one.

\subsection{Full kinematic reconstruction of the real emission}
\par So far, we have got the length of the tri-vectors $\vec{k}_n$ and $\vec{k}_{n+1}$. It is a standard kinematical problem to determine their directions in such a way that their sum $\vec{k}$ is parallel to $\vec{\overline{k}}_n$. We do not enter in further details about it.  
\par The last step is to calculate the $\beta$ parameter of the boost transformation $\Lambda$, eq.~\eqref{beta_boost}, and to boost ``back'' the other barred momenta in the real event
\begin{equation}
  k_i=\Lambda^{-1}\overline{k}_i, \quad i= 1,\cdots, n-1.
\end{equation}
The above mapping allows us to write the $(n+1)$-body phase space element in the factorized form  
\begin{equation}\label{fac_jacobian}
  \mathd\Phi_{n+1}= \mathd\Phi_{\text{rad}}\mathd\overline{\Phi}_n= J(\xi,y,\phi)\mathd\xi \mathd y \mathd\phi \mathd\overline{\Phi}_n,
\end{equation}
where we have expressed the radiation phase space in terms of the FKS variables with the jacobian function $J(\xi,y,\phi)$ taking into account the change of variables involved in the transformation. In order to extract the jacobian, we have to manipulate and compare the l.h.s and the r.h.s of eq.~\eqref{fac_jacobian}. Recalling eq.~\eqref{fac_recoil},
we perform the change of variables
\begin{equation}
  \vec{k}_n \to \vec{k}-\vec{k}_{n+1}
\end{equation}
in the three-body phase space, eq.~\eqref{three_body},
\begin{equation}
\begin{split}
    \mathd\Phi_3&=\frac{\mathd M^2_{\text{rec}}}{2\pi}\frac{\mathd^3\vec{k}}{2k^0_n(2\pi)^3}\frac{\mathd^3\vec{k}_{n+1}}{2k^0_{n+1}(2\pi)^3}\frac{\mathd^3\vec{k}_{\text{rec}}}{2k^0_{\text{rec}}(2\pi)^3}\\
    &\hspace{0.5cm}\times (2\pi)^4\delta^{(4)}(q-k-k_{\text{rec}})\,.
\end{split}
\end{equation}
In polar coordinates, we have
\begin{equation}
   \mathd^3\vec{k}= \underline{k}^2\mathd\underline{k}\mathd\Omega
\end{equation}
and, using as reference direction that of $\vec{k}$,
\begin{equation}
   \frac{\mathd^3\vec{k}_{n+1}}{2k^0_{n+1}(2\pi)^3} = \frac{q^2}{(4\pi)^3}\xi \mathd\xi \mathd\cos{\alpha}\mathd\phi\,,
\end{equation}
where $\alpha$ is the angle between $\vec{k}_{n+1}$ and $\vec{k}$ and $\phi$ is the azimuthal angle taking $\vec{k}$ as the reference direction. Hence
\begin{equation}\label{explicit_np1}
\begin{split}
 \mathd\Phi_{n+1} &= \frac{q^2}{(4\pi)^3}\xi \mathd\xi \mathd\cos{\alpha}\mathd\phi\frac{\underline{k}^2\mathd\underline{k}\mathd\Omega}{2k^0_n(2\pi)^3}\frac{\mathd M^2_{\text{rec}}}{2\pi}\\
 &\hspace{0.5cm}\times\frac{\mathd^3k_{\text{rec}}}{2k^0_{\text{rec}}(2\pi)^3}(2\pi)^4\delta^{(4)}(q-k-k_{\text{rec}})\mathd\Phi_{\text{rec}}.
\end{split}
\end{equation}
On the other hand, following the same arguments that led to eq.~\eqref{fac_recoil}, we can split the barred Born phase space into a two-body phase space and the phase space of the system recoiling against the emitting parton
\begin{equation}\label{split_born}
\begin{split}
  \mathd\overline{\Phi}_n&=\frac{\mathd\overline{M}^2_{\text{rec}}}{2\pi}\frac{\mathd^3\overline{\vec{k}}_n}{2\overline{k}^0_n(2\pi)^3}\frac{\mathd^3\overline{\vec{k}}_{\text{rec}}}{2\overline{k}^0_{\text{rec}}(2\pi)^3}\\
  &\hspace{0.5cm}\times(2\pi)^4\delta^{(4)}(q-\overline{k}_n-\overline{k}_{\text{rec}})\mathd\overline{\Phi}_{rec}.
\end{split}
\end{equation} 
Since $\overline k_n=q-\Lambda k_{\text{rec}}$, the delta function in eq.~\eqref{split_born} constrains the value of $\overline{k}_\text{rec}$ to be
\begin{equation}
  \overline{k}_{\text{rec}}= \Lambda k_{\text{rec}}.
\end{equation}
Then, exploiting the Lorentz invariance of the phase space element, we have
\begin{equation}\label{boost_invar}
\begin{split}
& \frac{\mathd M^2_{\text{rec}}}{2\pi}\frac{\mathd^3\vec{k}_{\text{rec}}}{2k^0_{\text{rec}}(2\pi)^3}(2\pi)^4\delta^{(4)}(q-k-k_{\text{rec}})\mathd\Phi_{\text{rec}}=\\
 &\phantom{aaaa}\frac{\mathd\overline{M}^2_{\text{rec}}}{2\pi}\frac{\mathd^3\overline{\vec{k}}_{\text{rec}}}{2\overline{k}^0_{\text{rec}}(2\pi)^3}(2\pi)^4\delta^{(4)}(q-\overline{k}_n-\overline{k}_{\text{rec}})\mathd\overline{\Phi}_{\rm rec},
\end{split}
\end{equation}
where the r.h.s and the l.h.s are related by the boost transformation $\Lambda$. In particular, we observe that
\begin{equation}
  \Lambda(q-k)= \Lambda k_{\text{rec}} = q-\overline{k}_n,
\end{equation}
so that the boost maps the argument of the delta function in the r.h.s into that of the delta
function in the l.h.s. Inserting eq.\eqref{explicit_np1} and eq.\eqref{split_born} into eq.\eqref{fac_jacobian} and using eq.\eqref{boost_invar}, we get
\begin{equation}\label{comp1}
  \begin{split}      
    & \frac{q^2}{(4\pi)^3}\xi \mathd\xi\, \mathd\cos{\alpha}\,\mathd\phi\frac{\underline{k}^2\mathd\underline{k}\mathd\Omega}{2k^0_n(2\pi)^3}=\\
    &\phantom{aaaaaaaaaaaaaaaaaaa}J(\xi,y,\phi)\mathd\xi\, \mathd y \,\mathd\phi\frac{\mathd^3\overline{k}_n}{2\overline{k}^0_n(2\pi)^3}.
  \end{split}
\end{equation}
By virtue of the mapping, the vectors $\vec{k}$ and $\vec{\overline{k}}_{n}$ are parallel so that in polar coordinates their angular elements are equal, $d\Omega=d\overline{\Omega}_{n}$. Then, from eq.~\eqref{comp1} we have
\begin{equation}
  \frac{q^2}{(4\pi)^3}\xi\frac{\underline{k}^2}{k_n^0}\mathd\cos{\alpha}\,\mathd\underline{k}= J(\xi,y,\phi) \frac{\overline{\underline{k}}_n^2}{\overline{k}_n^0}\,\mathd y \,\mathd\overline{\underline{k}}_n.
\end{equation}
and we are left with the computation of the jacobian of the two-variable-transformation
\begin{equation}
  J^{(2)} = \begin{vmatrix}
  \dfrac{\partial\overline{\underline{k}}_n}{\partial\underline{k}} & \dfrac{\partial y}{\partial\underline{k}}\\[1.0em]
  \dfrac{\partial\overline{\underline{k}}_n}{\partial\cos\alpha} & \dfrac{\partial y}{\partial\cos\alpha}
\end{vmatrix}
.
\end{equation}
This transformation is implicitly defined by the relations
\begin{equation}
\begin{split}
  \underline{k}_n&=\sqrt{\underline{k}^2+\underline{k}^2_{n+1}-2\underline{k}\,\underline{k}_{n+1}\cos\alpha}, \quad y=\frac{\underline{k}^2-\underline{k}_n^2-\underline{k}_{n+1}^2}{2\underline{k}_n\,\underline{k}_{n+1}},\\
  M^2_{\text{rec}}&=(q^0-k_n^0-\underline{k}_{n+1})^2-\underline{k}^2, \hspace{0.5cm} \overline{\underline{k}}_n=\frac{\lambda^{{1}/{2}}(q^2,M_{\text{rec}}^2,m^2)}{2q^0} ,
\end{split}
\end{equation}
where $\lambda$ is the kinematical Kallen function:
\begin{equation}
  \lambda(x,y,z)= x^2+y^2+z^2-2xy-2xz-2yz.
\end{equation} 
Applying the chain-rule for the derivative, it is straightforward to compute the jacobian. We get
\begin{equation}
  J^{(2)}= \frac{1}{\underline{k}_n^3}\frac{\underline{k}^2}{\overline{\underline{k}}_n}\frac{\overline{k}_n^0}{k_n^0}\big[k_n^0(\overline{k}_n^0-\underline{k}_{n+1})-m^2(1-{\underline{k}_{n+1}}/{q^0})\big]
\end{equation}
The final expression for the full jacobian $J$ is thus
\begin{equation}
\begin{split}\label{jacdef}
  J(\xi,y,\phi)&= \frac{q^2}{(4\pi)^3}\xi\frac{\underline{k}_n^3}{\overline{\underline{k}}_n}\frac{1}{k_n^0(\overline{k}_n^0-\underline{k}_{n+1})-m^2(1-{\underline{k}_{n+1}}/{q^0})}\\
  &=\frac{q^2}{(4\pi)^3}\xi\frac{\underline{k}_n^3}{\overline{\underline{k}}_n}\frac{2}{k_n^0(2\overline{k}_n^0-q^0\xi)-m^2(2-\xi)}
\end{split}
\end{equation}
Note that the denominator of $J$ vanishes in two regions:
\begin{itemize}
\item when approaching the curve $\xi=\xi(0)$ for $y>0$,
      behaving as $\xi(0)-\xi$
\item when approaching the curve $\xi=\xi(y)$, as
      $\sqrt{\xi(y)-\xi}$.
\end{itemize}
In the first case, the $k_n^3$ term in the numerator vanishes simultaneously 
as $(\xi(0)-\xi)^3$. It follows that the jacobian vanishes as
$J\sim(\xi(0)-\xi)^2$ for $\xi\to\xi(0)$ at fixed $y>0$. This result is 
coherent with what has been argued above regarding the degenerate points 
corresponding to the configuration with the emitter parton
at rest in the partonic centre-of-mass frame.
In the second region, the jacobian develops an integrable singularity, 
that can be dealt with by importance sampling techniques in Monte Carlo 
integration.
\subsection{Generation of radiation}
\blankout{
{\bf This section should be rewritten according to the latest implementation}
\begin{itemize}
\item Use of $k_t$ definition as in (54) (same as def)
\item Choice of upper bound incorporating a Jacobian
\item Jacobian singularity: treatment using the remnant trick.
\end{itemize}
}
The POWHEG master formula for the generation of radiation~\cite{Frixione:2007vw,Alioli:2010xd} is
\begin{equation}\label{POWHEG_master}
\begin{split}
  &d\sigma_\text{NLO} = \overline{B}(\Phi_n)d\Phi_n\bigg[\Delta_\text{NLO}(\Phi_n,t_{\rm min}) +\\
&\sum_\alpha\frac{\left[d\Phi_\text{rad}\Delta_\text{NLO}(\Phi_n,K_\perp(\Phi_{n+1}))R(\Phi_{n+1})\right]_{\alpha}^{\overline{\Phi}^\alpha_n=\Phi_n}}{B(\Phi_n)}\bigg],
\end{split}
\end{equation}
where $t_{\rm min}$ is an infrared cutoff, and
the NLO Sudakov form factor is given by 
\begin{equation}\label{NLO_SFF}
\begin{split}
  &\Delta_\text{NLO}(\Phi_n,p_T) =\theta(p_T-t_{\rm min}) \\
  &\exp{ \left[ -\sum_{\alpha}\int\frac{\left[d\Phi_\text{rad}R(\Phi_{n+1})\Theta(K_\perp(\Phi_{n+1})-p_T)\right]_{\alpha}^{\overline{\Phi}^\alpha_n=\Phi_n}}{B(\Phi_n)}\right]}. 
\end{split}
\end{equation}
In the case of a massless emitter, $K_\perp$ is a smooth function of the radiation 
variables, which is required to reduce 
to the transverse momentum in approaching the soft and collinear limits. 
For the massive case, in ref.~\cite{Barze:2012tt}
the following definition was proposed 
\begin{equation}\label{ktsq}
  K_\perp^2=2\frac{k^0}{p^0}p\cdot k = \frac{q^2}{2}\xi^2(1-\beta y_\text{phy}).
\end{equation}
$y_\text{phy}$ denotes the cosine of the physical angle between the emitter
and the emitted parton.\footnote{
  $y_\text{phy}$ must not be confused with the $y$ variable
  of the mapping. More specifically, in the region
  $\xi(0) \le \xi \le \xi_\text{max},\, y>0$ we have $y_\text{phy}=-y$,
  while in all the remaining region $y_\text{phy}=y$.}
Eq.~\eqref{ktsq} has the remarkable property of reducing continuously to the transverse
momentum in the massless limit. We assume it as our default scale choice.

According to the standard veto method, we look for a suitable upper
bound function $U$ of the integrand in the NLO Sudakov form factor, namely
\begin{equation}
    U(\xi,y)d\xi dy \ge \frac{R}{B} J(\xi,y) d\xi dy.
\end{equation}
For the sake of simplicity, we have omitted the integration on the azimuthal
angle $d\phi$, which results in a constant $2\pi$ factor.\\ 
We model the upper bound function on the asymptotic singular behavior of
the real matrix element near the soft and collinear singularities.
We recall that the jacobian of the mapping has a divergent behaviour near 
the curve $\xi=\xi(y)$. The upper bound function should have a behaviour not
weaker than the Jacobian near the singular regions, and
furthermore, it should be 
simple enough to allow us to perform an analytical integration in the 
constrained radiation phase space given by the cut $K_T^2>t$.\\ 
It is convenient to perform a change of integration variables from
$\xi,\,y$ to $\xi,\,K_T^2$.
Indeed, it turns out that $K_T^2$ is a monotonic decreasing function of
$y$ at fixed $\xi$, i.e. ${\partial K_T^2}/{\partial y} < 0$. \\
The inversion of this mapping is too complex to be performed analytically,\footnote{In fact,
  rather than proving analytically that  $K_T^2$ is a monotonic decreasing function of
  $y$ at fixed $\xi$, we demonstrated it numerically by checking it a large number of times
  for random values of the input parameters.}
but easy to perform numerically. We find that the 
associated jacobian ${\partial K_T^2}/{\partial y}$ has a
behaviour similar to that of the jacobian of the mapping $J$:
\begin{eqnarray}
&\sim&\frac{1}{\sqrt{\xi(y)-\xi}} \;\; \mbox{when}\;\; \xi \rightarrow \xi(y); \\ 
&\sim& (\xi(0)-\xi)^2  \;\; \mbox{when}\;\; \xi \rightarrow \xi(0)\;\;\mbox{for}\;\; y\geq0.
\end{eqnarray}
We now write
\begin{equation}\label{eq:U-Uprime}
  U = \frac{\partial K_T^2}{\partial y} U', 
\end{equation}
so that in the new integration variables the integrand becomes $U'$
\begin{equation}
  \int d\xi\, dy\, \Theta(K_T^2 -t) U =  \int d\xi dK_T^2 \Theta(K_T^2 -t) U'.
\end{equation}
$U'$ must have a simple form, and must have the appropriate behaviour to act as an upper bound
for the soft and collinear singularities of 
the real matrix element.
\subsubsection{Upper bound function}
The singular behaviour of the real matrix element squared is universal and
can be extracted in a straightforward manner by means of the
eikonal approximation. In terms of the radiation variables, we get
\begin{equation}
  \frac{R}{B} \sim \frac{N}{\xi^2(1-\beta y_\text{phy})} = \frac{N}{K_T^2},
\end{equation}
with $N$ a suitable normalization constant. On the other hand,
in the soft limit, the jacobian of the mapping behaves as
\begin{equation}
  J(\xi,y) \sim N'\xi.
\end{equation}
We must also take into account the behaviour in the soft limit of the 
jacobian term factorized in $U$:
\begin{equation}
  \frac{\partial K_T^2}{\partial y} \sim N''\xi^2.
\end{equation}
Putting all the three contributions together, we obtain the following
expression of the upper bound function $U'$
\begin{equation}\label{ubfun0}
  U'(\xi,K_T^2) = \frac{1}{K_T^2}\times\xi\times\frac{1}{\xi^2}=\frac{1}{\xi K_T^2}.
\end{equation}
A more complete analysis shows that mapping $J$ is enhanced (although not divergent) at large $\xi$ 
for $y\to -1$. In order to get a more efficient upper bound, we add the factor
$\frac{1}{1-{K_T^2}/{q^2}}$ to the previous expression. Hence, our final 
choice for the upper bound function $U'$ is
\begin{equation}\label{ubfun}
  U'(\xi,K_T^2) = \frac{1}{\xi K_T^2 (1-{K_T^2}/{q^2})}.
\end{equation}
\subsubsection{Integral of the upper bound function}
In order to integrate the upper bound function analytically, its domain
of integration has to be suitably enlarged. This can be done by interpreting
the $R/B$ expression as being defined in the larger domain, but as vanishing
outside of the physical domain. Since the veto procedure prescribes that a point
generated according to the upper bound function should be accepted with
a probability proportional to the value of the radiation function divided
by the upper bound function, points generated outside the physical domain
should always be vetoed according to the above interpretation.
From eq.~(\ref{ktsq}), we find the upper bound 
\begin{equation} \label{eq:tt2lim}
K_T^2 < K_\text{max}^2 \equiv \frac{q^2}{2}\xi_{\rm max}^2(1+\beta_0),
\end{equation}
where $\beta_0$ is the velocity of the emitter in the underlying Born
configuration (this follows from the fact that we always have $\beta \leq \beta_0$),
and we also find
\begin{equation} \label{eq:xilim}
  \frac{2K_T^2}{(1+\beta_0)} < \xi^2 < \frac{2K_T^2}{(1-\beta_0)},
\end{equation}
We thus take as our domain of integration the region in $K_T$ and
$\xi$ such that eqs.~(\ref{eq:tt2lim}) and (\ref{eq:xilim})
are satisfied. We notice that in this way $\xi$ can even become
larger than 1. In practice, however, adding also the $\xi<1$
or $\xi<\xi_{\rm max}$ limit would render the integration more
difficult, so we prefer to deal with it by vetoing. 
Defining
\begin{equation}
  \xi_{\,\stackrel{M}{m}{}}(K_t^2) \equiv \sqrt{\frac{2K_T^2}{q^2(1 \mp \beta_0)}},
\end{equation}
the integral of the upper bound function is then
\begin{equation}\label{eq:intupb}
  \begin{split}
    I(t) &= \int_t^{K_{\rm max}^2} \frac{dK_T^2}{K_T^2(1-{K_T^2}/{q^2})}\int_{\xi_m(K_t^2)}^{\xi_M(K_t^2)} \frac{d\xi}{\xi}\\
    & = \ln\left[\frac{K_{\rm max}^2}{q^2-K_{\rm max}^2}\frac{q^2-t}{t}\right]\,y_0,
    \end{split}
\end{equation}
where $y_0 \equiv (1/2)\ln[(1+\beta_0)/(1-\beta_0)]$ is the rapidity of 
the emitter in the underlying Born configuration.
Given a number $0<r<1$, the $t$ value generated
by solving the equation
$r=\exp[-2\pi N I(t)]$ is
\begin{equation}\label{gent}
  t = \frac{A}{1+A}q^2, \quad A=\frac{K_{\rm max}^2}{q^2-K_{\rm max}^2}\exp\left[\frac{\log{r}}{2\pi N y_0}\right].
\end{equation}

\subsubsection{Generation of radiation kinematics}
The algorithm for generating the radiation variables proceeds as follows:
\begin{enumerate}
\item We set the initial scale $t_0=K^2_{\rm max}$.
\item We generate a uniform random number
  \begin{displaymath} 0<r<\exp[-2\pi N I(t_0)], \end{displaymath}
  and get $t$ from eq.~(\ref{gent}). If $t$ is below $t_{\rm min}$,
  no radiation is generated, and the event is emitted as is.
\item We pick a new uniform random number $0<r'<1$ and we generate a value for 
$\xi$ as
  \begin{equation}
    \xi = \xi_\text{m}(t)\exp(y_0r').
  \end{equation}
  This is consistent with the distribution of $\xi$ at fixed $K_T^2$
  according to eq.~(\ref{eq:intupb}).
\item If $\xi>\xi_\text{max}$, we set $t_0=t$, and go back to the step 2.
\item If the veto condition is passed, given $t$ and $\xi$, we solve numerically for $y$ the implicit equation
  \begin{equation}
    K_T^2(\xi,y) = t.
  \end{equation}
  If a solution does not exist, we set $t=t_0$ and go back to step 2.
\item Now that $\xi$ and $y$ are available, we generate a random $\phi$, and
  compute the ratio $R=[R/B J(\xi,y)]/U(\xi,y)]$, with $U$ given in terms of $U'$ in eq.~(\ref{eq:U-Uprime}), and generate a new
    random number $0<r'''<1$. If $r'''>R$ we set $t_0=t$ and go back to the step 2.
    Otherwise, the event is accepted.
\end{enumerate}

\section{Phenomenology}\label{sec:pheno}
\subsection{comparison in the \bbfourl{} case}\label{sec:bb4l}
\begin{figure}
  \centering
 \includegraphics[width=0.45\textwidth]{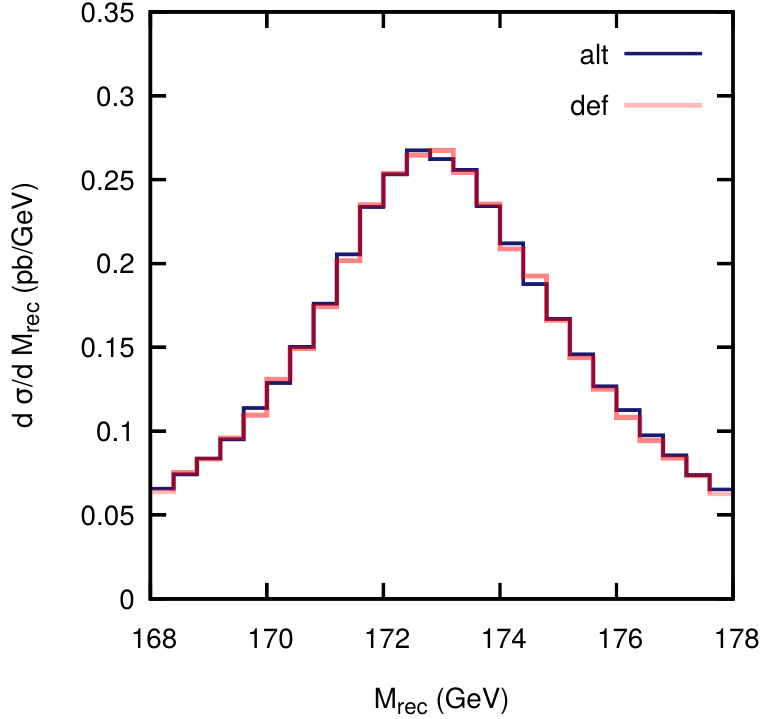}
 \caption{Invariant mass distribution of the reconstructed top quark
   mass, defined as the mass of the $W^+j_B$  or $W^-j_{\bar{B}}$ system,
   produced  with the \bbfourl{} generator, at the 8~TeV LHC. The
   two distributions are obtained with the default implementation
   of radiation from $b$ quarks (def), and with the new implementation
   presented here (alt).}\label{fig:topmass}
\end{figure}

\begin{figure}
  \centering
 \includegraphics[width=0.45\textwidth]{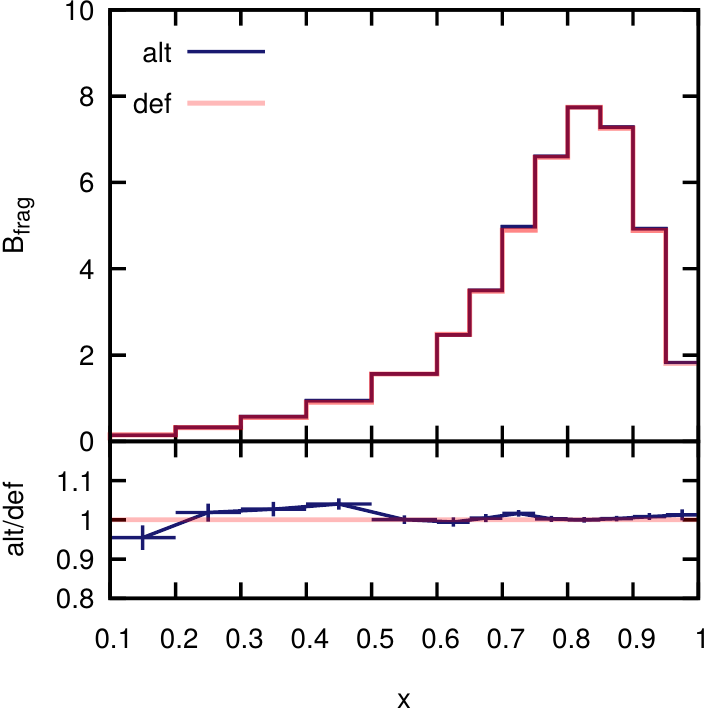}
  \caption{$B$ fragmentation function in top quark decay
    as defined in ref.~\cite{Jezo:2016ujg}, produced with the \bbfourl{} generator
    for the 8~TeV LHC. The default and alternative implementation of radiation from $b$ quarks are compared.}\label{fig:bfrag}
\end{figure}

\blankout{
We have verified that:
\begin{itemize}
\item The default default heavy quark treatment yields the same result
  as the \verb!alt_massive_map! one.
\item Bound violations in \verb!alt_massive_map! are not significantly worse
  than the  default.
\item The \verb!csidamp! option does not make a significant difference
  in the \verb!alt_massive_map! case.
\item In the \verb!alt_massive_map! case we have measured a generation
  speed of about 22 events per minute for our run. In the default
  case the generation speed is of about 8 events per minute.
\end{itemize}
We should collect the result from the intermediate run with \verb!alt_massive_map!,
and the corresponding result in the work with Oleari etal, and show very
few significant comparisons (they are essentially identical).
It may be useful to perform a new default run with the same parameters as the
\verb!alt_massive_map! one just for efficiency estimates.

Plots to show: reconstructed top mass peak at 8 TeV. B fragmentation function.
See if there is a jet profile that is worth showing.

\vskip 1cm
}
We have compared results obtained with the new method presented here, with those obtained
with the default \POWHEG{} settings for the \bbfourl{} generator of ref.~\cite{Jezo:2016ujg}.
We found remarkable agreement between the two results for all the distributions that we have
examined. Here we show only two of them, to convey the idea of the quality of the agreement.
These results were obtained for the 8TeV LHC collider, using the MSTW2008 PDF~\cite{Martin:2009iq}
set for reference only (other sets could be used as well~\cite{Nadolsky:2008zw,Ball:2014uwa}). 
In our simulations we make the $B$ hadrons stable.
Jets are reconstructed using the Fastjet~\cite{Cacciari:2011ma}
implementation of the anti-$k_{\rm\sss T}$ algorithm~\cite{Cacciari:2008gp}
with $R=0.5$.  We denote
as $B$~(${\bar B}$) the hardest (i.e.~largest \pT{}) $b$~($\bar{b}$)
flavoured hadron. The $B$ ($\bar{B}$) jet \bj{} (\bbj)
is defined to be the jet that contains the hardest
$B$ ($\bar{B}$). We discard events where the \bj{} and  \bbj{} coincide.
The hardest $e^+$ ($\mu^-$)
and the hardest $\nu_e$ ($\bar{\nu}_{\mu}$) are paired to reconstruct the $W^+$ ($W^-$).
The reconstructed top (antitop) quark is identified with the corresponding
$W^+j_B$ ($W^-j_{\bar{B}}$) pair.
We show the invariant mass of the $W-b$-jet system
(fig.~\ref{fig:topmass})
and the $B$ fragmentation function in top decay (Fig.~\ref{fig:bfrag}),
as defined in ref.~\cite{Jezo:2016ujg}, i.e. the
the $B$ energy in the reconstructed top rest frame normalized to the maximum value
that it can attain at the given top virtuality. In the curves, the \verb!alt! (for ``alternative'')
label stands for our new implementation, while \verb!def! (for ``default'') is the
current \POWHEG{} default.
As one can see, the agreement is very good. This also shows that details in the
implementation of radiation from the $b$ quark in top decays do not seem to have important
impact on physical observables.

We found that the efficiency and the generation rate of the new implementation 
are comparable with those of the \POWHEG{} default.

\subsection{$b$ production in hadronic collisions}
In this section we study the available \POWHEG{} implementations of radiation
from massive quarks for the \hvq{} generator~\cite{Frixione:2007nw}, i.e. the
default \POWHEG{} implementation and our new one. In spite of the fact
that the default formalism has been available for quite some
time~\cite{Barze:2012tt}, no such study has been performed so far.
We thus discuss it in this work, where we can also compare with
our new implementation.

\begin{figure}
  \centering
  \includegraphics[width=0.45\textwidth]{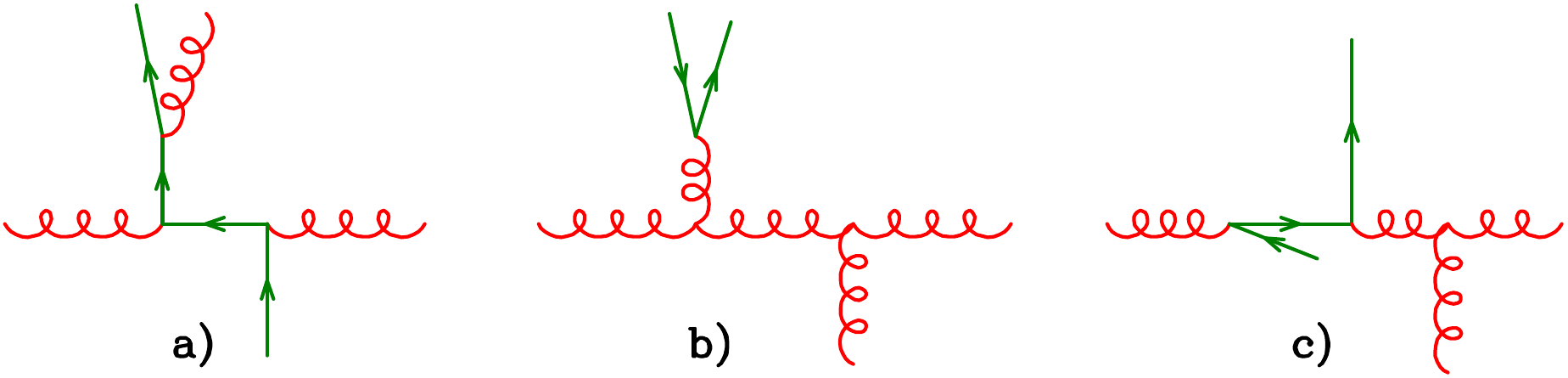}
  \caption{Example diagrams for the three mechanism that give rise to log-enhanced contributions
    in heavy flavour production: a) final state radiation from a quark; b) gluon
  splitting; c) flavour excitation.}\label{fig:logenhanced}
\end{figure}

The \hvq{} generator has been available for quite some time as a tool
to generate top, bottom and charm pairs in hadronic collisions. It is
designed to simulate correctly the production of a heavy flavour pair
when the logarithm of the ratio of the transverse momentum
of the heavy quark divided by its mass is not too large. 
This limitation arises because there are three
mechanisms, depicted in figure~\ref{fig:logenhanced},
involving radiation from the
final state quark, production of a heavy quark-antiquark pair via
final state gluon splitting and the splitting of an initial state gluon
into a heavy quark-antiquark pair (where one of the two quarks is scattered
at large transverse momentum), that can generate large logarithms involving
the mass of the heavy quark. In the inclusive cross section for the
production of a heavy quark with a given $p_T$, for example, they
generate logarithms of $p_T/m$ (see ref.~\cite{Nason:1989zy},
eq.~(5.1)).
The last two mechanisms are
commonly referred to as gluon splitting
and flavour excitation.
In spite of this, the \hvq{} generator has also been used to model relatively
large transverse momentum production of heavy flavours, as in
ref.~\cite{Cacciari:2012ny}.
There, the transverse momentum distribution of the heavy flavoured hadron
in \hvq{}  was compared with the more accurate
(but less exclusive) \FONLL{} prediction~\cite{Cacciari:1998it}.
It was found to be in rather good agreement. However, the
large uncertainties related to the non-perturbative fragmentation
of the heavy quark leads to the suspect that such agreement is
at least in part accidental.

\begin{figure}
  \centering
  \includegraphics[height=0.4\textwidth]{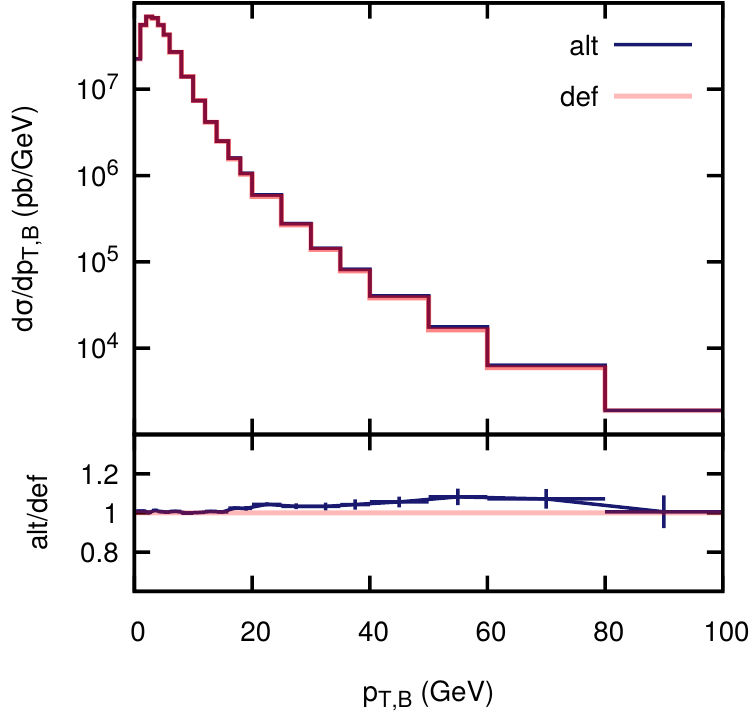}
  \caption{Comparison of \hvqaslalt{} and \hvqasldef{} for the
    transverse momentum distribution of the $B$ hadron
    at the 8TeV LHC.}\label{fig:hvq-Bpt-alt-def}
\end{figure}

\begin{figure}
  \centering  
  \includegraphics[height=0.4\textwidth]{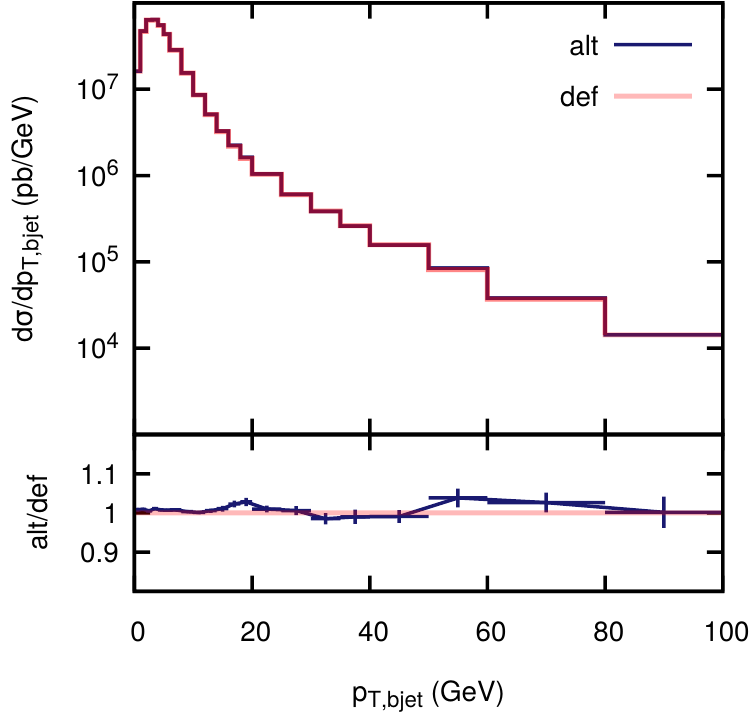}
  \caption{Comparison of \hvqaslalt{} and \hvqasldef{} for the
    transverse momentum distribution of the
  $b$-jet at the 8TeV LHC.}\label{fig:hvq-bjet-pt-alt-def}
\end{figure}

\begin{figure}
  \centering
  \includegraphics[width=0.4\textwidth]{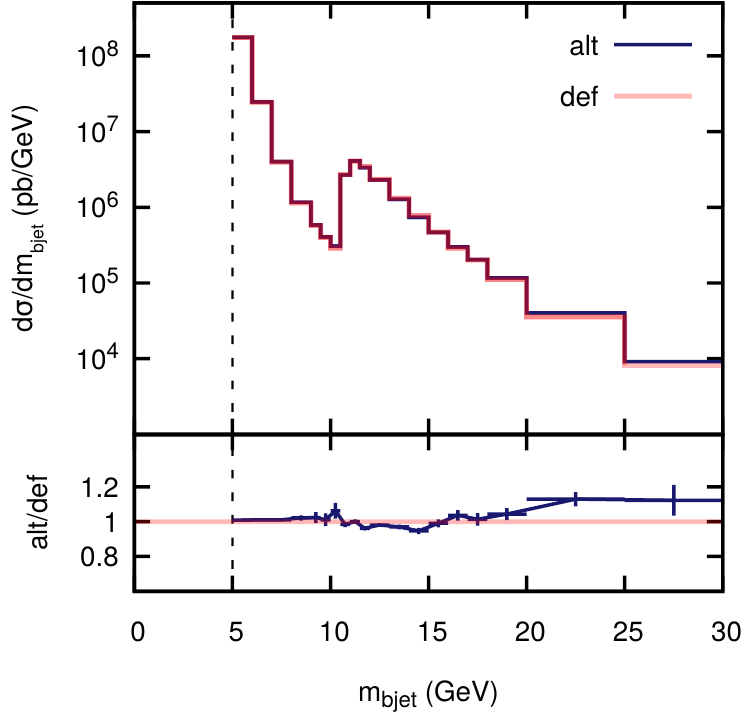}
  \caption{Comparison of \hvqaslalt{} and \hvqasldef{} for the
  $b$-jet mass.}\label{fig:hvq-bjet-mass-alt-def}
\end{figure}

We will now compare the results obtained with the default \hvq{}
generator, that we will label \hvqash{} (for ``no light'', meaning that
the heavy quark is treated as very heavy),
that treats as singular regions only the radiation from
massless partons (i.e. initial state radiation);
\hvq{} with the inclusion of the radiation from the heavy
quark as a singular region will be labeled \hvqasl{}
(for ``as light'', meaning that
the heavy quark is treated as a light parton). Furthermore, the default treatment
of the heavy quark radiation region will be denoted as \hvqasldef{},
while  the new implementation presented here will be called \hvqaslalt{}.
In figs.~\ref{fig:hvq-Bpt-alt-def}, \ref{fig:hvq-bjet-pt-alt-def} and~\ref{fig:hvq-bjet-mass-alt-def}
we show a comparison of \hvqasldef{} and \hvqaslalt{}.
We can immediately see that we do not find important differences
between the two methods, consistently with what was found in the
\bbfourl{} case.
The settings are similar to the \bbfourl{} case:
we make the $B$ hadrons stable, and define the
$b$ (${\bar b}$) jets as the jets containing the hardest $b$ (${\bar b}$)
flavoured hadron, with the jets defined as in the \bbfourl{} case.
However, we do not exclude the case when both hardest $b$-flavoured
hadrons are in the same jet.
We perform the calculation for the LHC at 8 TeV, using {\tt NNPDF30\_nlo\_as\_0118}
pdf set~\cite{Ball:2014uwa}.
As one can see, the two implementations are in excellent agreement.
Observe the jump at 10 GeV in the \bj{} mass. It is due to the case in which
the $b$ and ${\bar b}$ flavoured hadrons are both in the jet cone.
From figure~\ref{fig:hvq-bjet-mass-alt-def} we also see that for jet
masses above 10~GeV the gluon splitting configuration dominates.

We found that the new implementation has a generation efficiency, which is 
estimated from the numbers of vetoes in FSR generation, three times greater
than the default one. This leads to a generation rate of 1316 events per minute, 
against the 298 events per minute of the \POWHEG{} default, which corresponds
to a gain more than a factor of 4.

\begin{figure}
  \centering
  \includegraphics[height=0.4\textwidth]{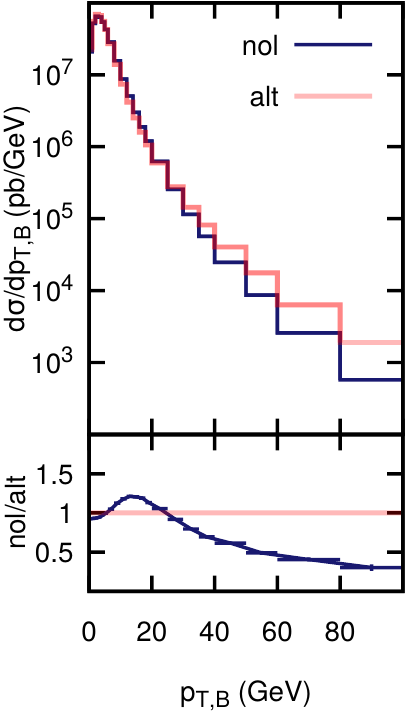}
  \includegraphics[height=0.4\textwidth]{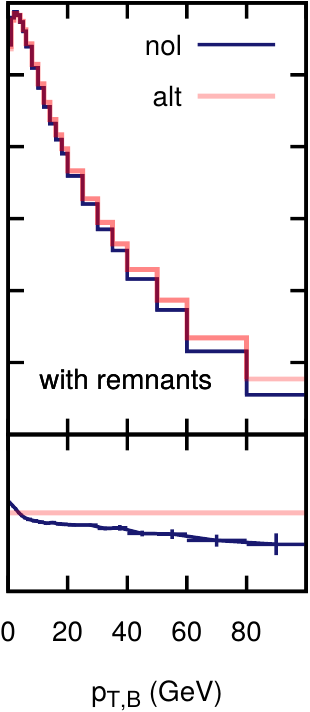}
  \caption{Left panel: comparison of \hvqaslalt{} and \hvqash{} for the
    transverse momentum distribution of the $B$ hadron 8TeV LHC.
    Right panel: same comparison with the treatment of the enhanced regions
    using remnants, as discussed in the text. }\label{fig:hvq-Bpt-alt-nol}
\end{figure}

\begin{figure}
  \centering
  \includegraphics[height=0.4\textwidth]{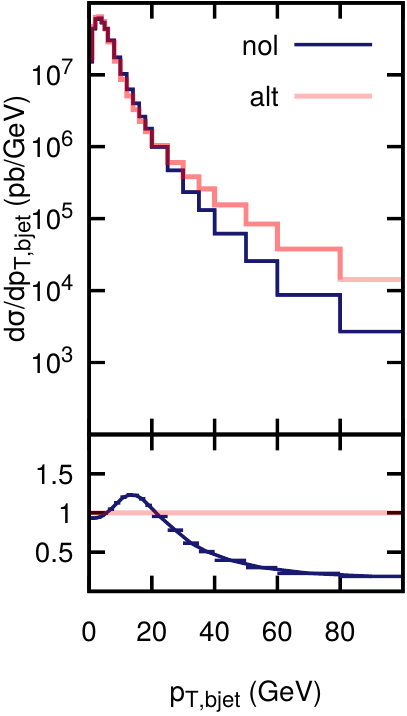}
  \includegraphics[height=0.4\textwidth]{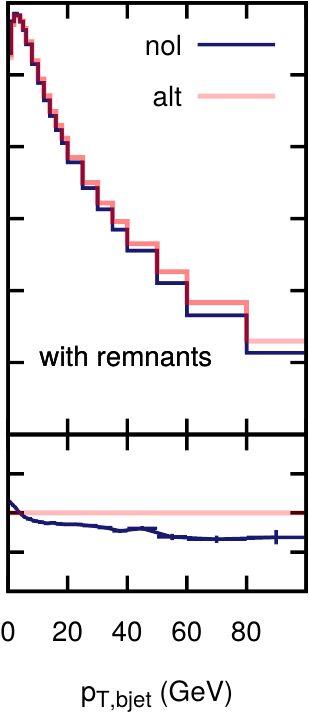}
  \caption{Same as in fig.~\ref{fig:hvq-Bpt-alt-nol} for the $p_T$
     of the $b$-jet.}
    \label{fig:hvq-bjet-pt-alt-nol}
\end{figure}

\begin{figure}
  \centering
  \includegraphics[height=0.4\textwidth]{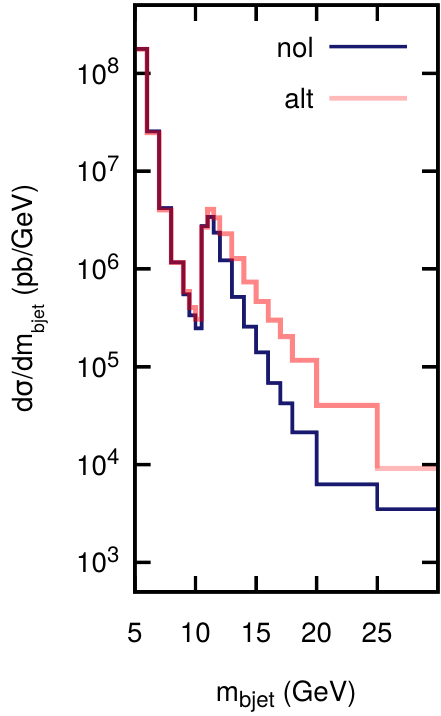}
  \hskip -0.3cm
  \includegraphics[height=0.4\textwidth]{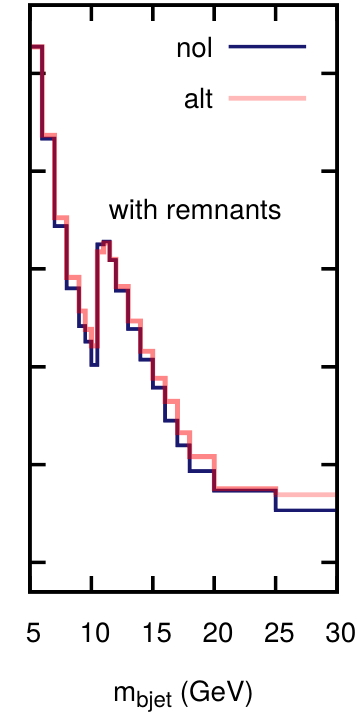}
  \caption{Same as in fig.~\ref{fig:hvq-Bpt-alt-nol} for the
  $b$-jet mass.}\label{fig:hvq-bjet-mass-alt-nol}
\end{figure}

We now show in the left panels of Figs.~\ref{fig:hvq-Bpt-alt-nol},
\ref{fig:hvq-bjet-pt-alt-nol} and~\ref{fig:hvq-bjet-mass-alt-nol}
the comparison among the \hvqaslalt{} and \hvqash{}. Here we see considerable
differences, especially in the large-momentum tail of the $B$ and \bj{}
transverse momentum distribution, the \hvqaslalt{} ones being much harder.
The mass of the $b$~jet is also remarkably different. The large difference above
10 GeV hints to the fact that heavy quark pair production via the splitting
of a large transverse momentum gluon is treated in a very different way in the
two cases, and that this difference may be the cause of the large discrepancy
in the transverse momentum distribution of the $b$ hadron.

The difference between the \hvqaslalt{} and \hvqash{} cases should not
come as a surprise.
The generation of radiation is performed in the \hvqash{} case
according to the formula
\begin{eqnarray}\label{eq:basic_hvq}
  \mathd \sigma &=& \mathd \Phi_B \tilde{B}(\Phi_B) \exp\left[\int
    \frac{R(\Phi_B,\Phi'_{\rm rad})}{B(\Phi_B)} \theta(k_t'-k_t)
    \mathd \Phi'_{\rm rad}\right] \nonumber \\
& \times &   \frac{R(\Phi_B,\Phi_{\rm rad})}{B(\Phi_B)}  \mathd\Phi_{\rm rad}\,,
\end{eqnarray}
where $k_t$ is the transverse momentum of the emitted gluon with respect to
the beam axis, since the only singular regions that are considered
there are the initial-state radiation (ISR) ones. The strong coupling constant and
the parton densities are evaluated by default at a scale equal to the transverse
mass of the heavy quark at the level of the underlying Born kinematics
\begin{equation} \label{eq:hvqscale}
\mu_f=\mu_r=\sqrt{k_{t,q}^2+m_q^2}
\end{equation}
in the $\tilde{B}$ function, while they are evaluated at a scale $k_t$ (or $k'_t$)
in the $R/B$ ratios appearing in formula~(\ref{eq:hvqscale}). Since $\tilde{B}$ and $B$ are
of order $\as^2$, while $R$ is of order $\as^3$, this means that in practice
two powers of the strong coupling are evaluated at the scale of
eq.~(\ref{eq:hvqscale}), while one power is evaluated at a scale $k_t$.
The mismatch in the scale used in $\tilde{B}$ and in the $B$ appearing in the
ratios, combined with the exponential, leads as usual to the correct Sudakov
form factor for initial state emission.
\subsubsection{Problematic regions}\label{sec:problematic}
In case the transverse momentum of the gluon is small, the scale
assignments and the Sudakov form factor describe the process appropriately.
It can happen however, that the real emission kinematics is near the gluon splitting, flavour excitation or quark radiation regimes. In these cases the gluon transverse momentum is
not small. Furthermore, the numerator $R$ in the integrand may be
enhanced with respect to the denominator, thus yielding a damping of the real
cross section that is not justified.
Also the scale choices are not appropriate.
For example, in the case of production of a high transverse momentum heavy quark pair
according to the gluon splitting mechanism, the appropriate scale should correspond
to two powers of $\as$ evaluated at the gluon transverse momentum, and one power of $\as$
evaluated at the scale of the order of the invariant mass of the heavy quark pair.

The adoption of the methods illustrated in ref.~\cite{Barze:2012tt} and in the
present work for dealing with radiation from a heavy quark leads
to the correct treatment of the radiation from the heavy, quark provided
all remaining regions are treated correctly. This is in fact what happens
in the case of the \bbfourl{} generator, where there is only one enhanced region,
but it is not the case for the \hvqasl{} generator,
that does not treat in a proper way the two regions of gluon splitting and %
flavour excitation.
Thus, the \hvqash{} and the \hvqasl{} generators will end
up treating the enhanced regions in different (and in both cases
incorrect) ways. In fact, while
in the \hvqash{} case the enhanced regions will all be treated as if
they were ISR processes, in the \hvqasl{} case
they will be split, and treated in part as ISR processes, and in part
as radiation from the heavy quarks.
In order to test this hypothesis, and in order to explore possible
strategies to deal with this problem, we proceed as follows.
It is possible in \POWHEG{} to further separate out the real cross section into
two terms, such that only one term has singular behaviour, while the remaining term,
being finite, can be integrated independently. In the \hvq{} case, this means
\begin{equation}
  R=R^{(s)}+R^{(r)}\,.
\end{equation}
Eq.~(\ref{eq:basic_hvq}) is then replaced by
\begin{eqnarray}\label{eq:rsrmsudakov}
 && \mathd \sigma = \mathd \Phi_B \tilde{B}^{(s)}(\Phi_B) \exp\left[\int
    \frac{R^{(s)}(\Phi_B,\Phi'_{\rm rad})}{B(\Phi_B)} \theta(k_t'-k_t)
    \mathd \Phi'_{\rm rad}\right] \nonumber \\
  && \times  \frac{R^{(s)}(\Phi_B,\Phi_{\rm rad})}{B(\Phi_B)}  \mathd\Phi_{\rm rad}
  +\int \mathd \Phi_B \mathd\Phi_{\rm rad} R^{(r)}(\Phi_B,\Phi_{\rm rad})\,.
\end{eqnarray}
We can exploit this mechanism in order to separate out the enhanced regions,
in such a way that we can treat them in a more uniform way with our generators.
In particular, we separate out the gluon splitting and flavour excitation
processes in all cases. In the \hvqash{} case we also separate out
the regions of radiation from the heavy quarks, in such a way that they are treated in
a more transparent way. Observe that in performing this separation we rely upon the
fact that the three enhanced region are not really singular, since the quark mass
cuts off the collinear singularities, and thus the remnant term is actually finite.

We define the distance of a real configuration from a given
enhanced region as follows
\begin{equation}
\begin{array}{lll}
  d_{\rm isr}  &= k_t^2, \phantom{aaaaaaaaaaaaaa}  d_{\rm glsp} &=
  2 k_q\cdot k_{\bar q}\frac{k_q^0 k_{\bar q}^0}{(k_q^0 + k_{\bar q}^0)^2}, \\
  d_q&=2 k_q\cdot k\frac{k^0}{k_q^0}+m_q^2,
  \hfill\hfill  d_{\bar q}&=2 k_{\bar q}\cdot k\frac{k^0}{k_{\bar q}^0}+m_q^2, \\
  d_{q,{\rm flex}} &= k_{{\bar q},\perp}^2 + m_q^2,
  \hfill\hfill d_{{\bar q},{\rm flex}} &= k_{q,\perp}^2 + m_q^2,
  \end{array}
\end{equation}
where in the first line the distances for ISR and gluon splitting are given,
in the second line those for radiation from the heavy quarks, and in the last
line the ones for flavour excitation. We then define, for the
\hvqash{} generator
\begin{eqnarray}
  D &=& \frac{d_{\rm isr}^{-1}}{d_{\rm isr}^{-1}+d_{\rm glsp}^{-1}+ d_q^{-1}+d_{\bar q}^{-1}
    +d_{q,{\rm flex}}^{-1}+d_{{\bar q},{\rm flex}}^{-1} }, \nonumber \\
  R^{(s)}&=&R D,\qquad R^{(r)}=R (1-D)\,. \label{eq:damp1}
\end{eqnarray}

For the \hvqaslalt{} and \hvqasldef{} generators, we define
\begin{eqnarray}
  D &=& \frac{d_{\rm isr}^{-1}+ d_q^{-1}+d_{\bar q}^{-1}}{d_{\rm isr}^{-1}+d_{\rm glsp}^{-1}+ d_q^{-1}+d_{\bar q}^{-1}
    +d_{q,{\rm flex}}^{-1}+d_{{\bar q},{\rm flex}}^{-1} } \\
  R_i^{(s)}&=&R_i D,\qquad R_i^{(r)}=R_i (1-D)\,,
\end{eqnarray}
where the index $i$ labels the three singular regions that \POWHEG{} is handling. In this
case, the cross section is damped if the kinematics is near a singular region that
is nether ISR nor FSR, i.e. only gluon splitting and flavour excitation kinematics
are separated into the $(r)$ component.

There is one more issue that needs to be considered when using a damping factor in
\POWHEG{}. By default, when evaluating the $R^{(r)}$ component (called ``real remnant''),
the scale choice is the same as for $\tilde{B}$, i.e. it is eq.~(\ref{eq:hvqscale})
applied to the underlying Born kinematics, that depends upon the considered singular
region. This would lead to a different scale choice for the remnants in \hvqash{} and \hvqasl{}.
In order to avoid that, we should set the scale on the basis
of the real kinematics. This can be done in \POWHEG{} by setting appropriate flags and
by modifying the code that computes the scales for the process. Our scale choice is
\begin{equation}\label{eq:remscalechoice}
  \mu_f=\mu_r=\frac{1}{2}\left[\sqrt{k_{t,q}^2+m_q^2}+\sqrt{k_{t,{\bar q}}^2+m_q^2}+k_t\right]\,,
\end{equation}
that has the correct limit to the underlying Born scale both in the ISR and in the FSR
case.

The result of this procedure is shown
in the right panels of Figs.~\ref{fig:hvq-Bpt-alt-nol}, \ref{fig:hvq-bjet-pt-alt-nol} and~\ref{fig:hvq-bjet-mass-alt-nol}.
We notice a remarkable improvement in
the agreement, although some important differences do remain. This is not
unexpected, since in the two cases radiation from the heavy quark is treated
in a very different way. It is interesting to notice that the $B$ and the $\bj$
spectra computed with the \hvqash{} {\emph without} remnants (which is the default
in the standard \hvq{} generator), is in fair agreement with the \hvqaslalt{} one
when the enhanced regions are separated using the remnants. Since the
default \hvq{} program gives a description of the transverse momentum distribution
of $B$ hadrons that is in fair agreement with the FONLL calculation, we infer that
also the \hvqaslalt{} prediction will display a similar agreement, provided the
gluon splitting and flavour excitation region are treated separately as remnants.

The \hvqaslalt{} (or equivalently the \hvqasldef{} generator), with the remnant
separation discussed above, seems to be at this point the generator that may give the best
description of $b$ production data at hadron collider. We should not forget, however,
that some flexibility still remains in the treatment
of the remnant (in this work we have made a definite scale choice for the remnants
in order to have a clearer comparison
with the \hvqash{} generator).
We also notice from figs.~\ref{fig:hvq-Bpt-alt-nol} and \ref{fig:hvq-bjet-pt-alt-nol} that
after the remnants are introduced, the $B$-hadron and $b$-jet $p_T$ spectra become
softer. This seems to be in contrast with the discussion at the beginning of sec.~\ref{sec:problematic}.
On the other hand, this result may be due to the particular scale choice that we have performed
for the real graphs, and that \POWHEG{} applies automatically also to the remnants.
This scale turns out to be higher than the typical scale involved in the region
discussed at the beginning of sec.~\ref{sec:problematic}. A better approach would be
to introduce the possibility of alternative scale choices in the remnants, including the possibility
of performing a different scale choice depending upon which enhanced region one is considering.
We refrain here from carrying out such study, since we believe that
comparison with data on single inclusive $b$-hadron and $b$-jet production
(see ref.~\cite{Khachatryan:2011mk,Aad:2012jga,Acosta:2004yw} and references therein)
and on correlations of $b\bar{b}$ pairs~\cite{Abbott:1999se,Khachatryan:2011wq},
would be needed in order to make progress in this direction, and this is beyond the scope
of the present work.
Such study would however be very valuable, and not only for the purpose of testing QCD
in bottom production. The production of top pairs is one of the most important background
process at the LHC, including also its future high luminosity and eventually high energy developments.
An accurate simulation of the $t\bar{t}$ background would be very valuable
for the LHC experimental collaborations, and it is quite clear that a model of $b$ production
yielding a good description of the data from low to high transverse momenta
should be also well suited to describe top production in all the needed phase space.
Furthermore, these studies could lead to an improved description of top production at
large transverse momentum, that, as reported
currently by CMS and ATLAS
(see \cite{Sirunyan:2017mzl,Aaboud:2016iot} and references therein) 
 seems to be not well described by theoretical models.

\section{Conclusions} \label{sec:conc}
In this paper we have presented a method for implementing radiation from a heavy quark in
the \POWHEG{} framework. This method is considerably simpler and more transparent than
the one presented in ref.~\cite{Barze:2012tt}, and it has a much better numerical performance.
The present method overcomes a problem related to the fact that the most natural map
from an underlying Born configuration and a set of FKS-like radiation variables for radiation
from a massive quark
does not have a unique inverse in the whole kinematic region. The \POWHEG{} inverse map
has thus two solutions in some region of phase space,
and in the present work it is shown how, by giving up the physical connection
of the $y$ variable to the gluon emission angle in a very limited region of phase space one
can pick one of the two solutions in such a way that we still have a single valued inverse mapping.
We have examined the output of the new method in the framework of the generators of ref.~\cite{Jezo:2016ujg}
and~\cite{Frixione:2007nw}.
We found that the new method yields results that are very consistent with the previous
one, that is at this moment the \POWHEG{} default. This is reassuring, since it shows that
details of the implementation do not impact in a visible way the physics result, and also
it shows that the previous implementation, in spite of being quite contrived,
is in fact correct.

In this work we have also examined for the first time the impact of the inclusion of the
singular region associated with radiation from the heavy quark
in the case of the \hvq{} generator. We have shown that, unless one separates the enhanced
gluon splitting and flavour excitation contribution from the real contribution that
are dealt with by the \POWHEG{} radiation formula, and treats them as remnants, one finds
results that are in considerable disagreement with the traditional \hvq{} implementation.
On the other hand, it seems that performing this separation is the appropriate thing to
do for a consistent
modeling of the process. We notice that such modeling would be of great interest also for
its potential application to top pair production, that is a very important background to
many LHC physics studies.

\begin{acknowledgements}
We thank Matteo Cacciari for useful exchanges. The work of L.B. and F.T. has been
supported in part by the Italian Ministry of Education and Research MIUR, under project n$\rm{^o}$~2015P5SBHT.
\end{acknowledgements}

\bibliography{paper}

\end{document}